\documentclass[12pt]{article}
\usepackage{a4wide}
\usepackage{authblk}
\usepackage{hyperref}
\usepackage[top=25truemm,bottom=25truemm,left=15truemm,right=15truemm]{geometry}
\usepackage{setspace}
\usepackage{amsmath,amssymb,mathrsfs,amsbsy,latexsym,amsfonts,amsthm}
\usepackage{physics}
\usepackage{fancyhdr}
\usepackage[Symbol]{upgreek}

\usepackage{graphicx}           

\usepackage{color}
\usepackage{slashed}
\usepackage{tensor}

\newcommand{\red}[1]{{\color{black}#1}}

\newcommand{\al}[1]{\begin{align}#1\end{align}}

\newcommand{\paren}[1]{\left(#1\right)}

\newcommand{\sqbr}[1]{\left[#1\right]}
\newcommand{\br}[1]{\left\{#1\right\}}

\newcommand{\nn}{\nonumber\\}

\newcommand{\p}{\partial}
\newcommand{\Slash}[1]{{\ooalign{\hfil/\hfil\crcr$#1$}}} 
\newcommand{\outst}{\tensor[_0]{\bra{\text{out}}}{}}
\newcommand{\inst}{\ket{\text{in}}_0}

\usepackage{epsf}

\makeatletter

\@addtoreset{equation}{section}
\makeatother

\begin{document}
\title{
\bf \Large 
Notes on the gravitational, electromagnetic\\ and axion memory effects
\vskip 0.5cm
}
\author[a,b]{
Yuta~Hamada\thanks{\tt yhamada@wisc.edu}
}
\author[c]{
Sotaro~Sugishita\thanks{\tt sugishita@het.phys.sci.osaka-u.ac.jp}
\vspace{5mm}
}
\affil[a]{\it\normalsize Department of Physics, University of Wisconsin-Madison, Madison, WI 53706, USA\\
\vspace{1mm}}
\affil[b]{\it\normalsize KEK Theory Center, IPNS, KEK, Tsukuba, Ibaraki 305-0801, Japan\\
\vspace{1mm}}
\affil[c]{\it\normalsize Department of Physics, Osaka University, Toyonaka, Osaka, 560-0043, Japan}
\setcounter{Maxaffil}{0}
\date{}


\maketitle
\thispagestyle{fancy}
\renewcommand{\headrulewidth}{0pt}

\abstract{\noindent \normalsize
We investigate the memory effects associated with the kicks of particles. \red{Recently, the equivalence between the memory effect and soft theorem has been established. By computing the memory effect from the radiation solutions,} we explicitly \red{confirm} that, in addition to the leading piece, the subleading and subsubleading soft theorems are equivalent to the subleading and subsubleading memory effects, respectively. It is known that the memory effects can be probed by the displacements or kicks of the test particles. We point out that the these memory effects are also probed by the permanent change of the direction of the spin. We also show that the axion memory effect, recently proposed by the current authors,  can be detected as the change of the spin of the test particle.
We discuss that if we consider the magnetic monopole as an external particle, the parity-odd electromagnetic memory appears.
}

\newpage
\normalsize
\section{Introduction}\label{sec:intro}
The detection of the gravitational wave~\cite{Abbott:2016blz} \red{has opened} up the era of gravitational astronomy and cosmology.
It is known that, in addition to the oscillatory gravitational wave, there exists the non-oscillatory contribution~\cite{Zeldovich1974,Braginsky:1986ia,Braginsky1987,Christodoulou:1991cr}, which is called gravitational memory effect.
This results in the permanent displacement of the proper distance between the free falling detector particles.
The understanding of the memory effect is quite important for the future gravitational wave experiment.
Recently, the memory effect in cosmological setup \red{has been}  investigated~\cite{Bieri:2015jwa,Chu:2015yua,Kehagias:2016zry,Chu:2016qxp,Tolish:2016ggo,Hamada:2017gdg,Bieri:2017vni}.
Other than the experimental aspect, the memory effect is theoretically interesting because it has a strong connection to the notion such as the soft graviton theorem and asymptotic symmetry~\cite{Strominger:2014pwa,Hamada:2017uot,Campiglia:2017dpg,Hamada:2017bgi,Campiglia:2017xkp,Hamada:2018vrw,Li:2018gnc}.

Inspired by the gravitational memory effect, an electromagnetic analog of the gravitational memory effect \red{was} discovered~\cite{Bieri:2013hqa}, where it \red{was} shown that the permanent kick of the momentum is observed instead of the permanent displacement. More recently, the axion memory effect \red{was} also proposed by the present authors~\cite{Hamada:2017atr}.

In this paper, we investigate the gravitational, electromagnetic and axion memory effects. 
We first revisit the derivation of the memory effect induced by the kick of the particle which sources the graviton, photon or axion. 
\red{It has been known for a long time \cite{Yennie:1961ad} that the classical radiation by the kick is related to the soft factor appeared in the leading soft theorem. 
	Recently, this relation has been revisited with the connection to memory effects \cite{Strominger:2014pwa, Pasterski:2015tva, Mao:2017wvx}.
	We explicitly compute the radiation by the kick of particles and confirm that the soft factor is nothing but the leading memory of the kick, which is characterized by the step function $\Theta(u)$ term in the radiation.  In addition to the leading order, we look at the subleading and the subsubleading memories characterized by the $\delta(u)$ and $\delta'(u)$ terms in the classical radiation, and show that they are related to the soft factors in the subleading and subsubleading soft theorems.}
Then, we discuss the decomposition of the E and B modes\footnote{These correspond to the parity even and odd modes, respectively.} of the memory effect. 
Compared with the earlier studies, we \red{also} find the following new things. 
\red{First}, we derive the subleading order axion memory effect which is induced by the kick of the particle coupled with the axion. 
\red{Second}, it is found that the electromagnetic leading memory contains the B-mode if we allow the magnetic monopole as an external state.\footnote{\red{The soft photon theorem in the presence of the magnetic monopole is known~\cite{Strominger:2015bla}. On the other hand, the memory effect induced by the monopole was not studied before. Previous studies on the electromagnetic memory effect~\cite{Bieri:2013hqa,Winicour:2014ska,Susskind:2015hpa,Mao:2017axa} focused on the source without magnetic charges.} }
\red{Third}, we show that the subsubleading memory in gravity cannot be observed if one can observe only the trajectory and spin of the test particle at far past and future. Finally, we point out that the (sub)leading memory effect of gravity, photon and axion can be detected  as a permanent change of the spin direction.

This paper is organized as follows. In Sec.~\ref{Sec:memory_effect}, we derive the memory effects by solving the classical equations of motion using the Green function. The detection of the memory effect is discussed in Sec.~\ref{Sec:detection_memory}. The summary is presented in Sec.~\ref{Sec:summary}. The definition of the $E$ and $B$ decomposition is collected in App.~\ref{App:EB_decomposition}. In App.~\ref{App:soft_pion_theorem_spin1}, the proof of the soft pion theorem for spin $1$ particle is presented. We also argue that the spin $2$ particle does not contribute to the soft pion theorem at the subleading order.

\section{Derivation of the memory effects}\label{Sec:memory_effect}

\subsection{Photon}
\subsubsection{Derivation of the electromagnetic memory effect}
We consider the electromagnetic potential induced from kicks of charged particles, and see that it leads to the electromagnetic memory. \red{See Refs.~\cite{Bieri:2013hqa,Winicour:2014ska,Susskind:2015hpa,Mao:2017axa} for studies on the electromagnetic memory effect.}
\red{As mentioned in section~\ref{sec:intro}, it has been known that classical radiation is related to the soft factors in the soft theorem \cite{Yennie:1961ad, Peskin:1995ev, Mao:2017wvx}. 
We see that the soft factors characterize step function terms and delta function terms in the classical radiation [see \eqref{eq:em_rad_memory}].}

The motion of charged particles is as follows:  
The particles first move as 
\begin{align}
y_n^\mu(\tau)=
\frac{p^\mu_n}{m_n}\tau +x_0^\mu  \,\quad (\tau<0)\,, 
\label{initial_tr}
\end{align}
with charges $e_n$, and after a kick, they move as\footnote{The number of particles can be changed.}  
\begin{align}
y_{n'}^\mu(\tau)=
\frac{p^\mu_{n'}}{m_{n'}}\tau +x_0^\mu  \,\quad (\tau>0)\,, 
\label{final_tr}
\end{align}
with charges $e_{n'}$. 
$x_0^\mu$ is a spacetime point $x_0^\mu=(t_0, \vec{x}_0)$ where the kick occurs. 
We assume that charges are conserved 
\begin{align}
\sum_n e_n = \sum_{n'} e_{n'}\,.
\end{align}
We also represent the momenta as  
\begin{align}
p_n^\mu = \omega_n (1,\vec{v}_n)\,, \quad p_{n'}^\mu = \omega_{n'} (1,\vec{v}_{n'})\,.
\end{align}
They satisfy\footnote{Signature is $(-,+,+,+)$.} 
\begin{align}
p_n^2= -\omega_n^2 (1-v_n^2)=-m_n^2\,, \quad p_{n'}^2= -\omega_{n'}^2 (1-v_{n'}^2)=-m_{n'}^2\,.
\end{align} 
The electromagnetic current is given by 
\begin{align}\label{Eq:elemag_source}
j^\mu(x)=\sum_n \int^0_{-\infty} d \tau\, \frac{e_n p_n^\mu}{m_n}\, \delta^{(4)}(x-y_n(\tau))
+\sum_{n'} \int^{\infty}_0 d \tau\, \frac{e_{n'} p_{n'}^\mu}{m_{n'}}\, \delta^{(4)}(x-y_{n'}(\tau))
\end{align}

We solve the Maxwell equation in the Lorenz gauge $(\partial_\mu A^\mu=0)$,  
\begin{align}
- \partial^2 A^\mu = j^\mu
\end{align} 
The retarded solution is given by 
\begin{align}
&A^\mu(x)\nn
&=\Theta(|\vec{x}-\vec{x}_0|-t+t_0)\sum_n \frac{e_n p_n^\mu}{4\pi m_n\sqrt{|\vec{x}-\vec{x}_0|^2-\paren{\frac{\vec{v}_n\cdot(\vec{x}-\vec{x}_0)}{v_n}}^2+\frac{1}{1-v_n^2}\paren{\frac{\vec{v}_n\cdot(\vec{x}-\vec{x}_0)}{v_n}-v_n (t-t_0)}^2}}
\nn
&+\Theta(-|\vec{x}-\vec{x}_0|+t-t_0)\sum_{n'} \frac{e_{n'} p_{n'}^\mu}{4\pi m_{n'}\sqrt{|\vec{x}-\vec{x}_0|^2-\paren{\frac{\vec{v}_{n'}\cdot(\vec{x}-\vec{x}_0)}{v_{n'}}}^2+\frac{1}{1-v_{n'}^2}\paren{\frac{\vec{v}_{n'}\cdot(\vec{x}-\vec{x}_0)}{v_{n'}}-v_{n'} (t-t_0)}^2}}
\,.
\label{eq:em_ret_sol}
\end{align}
The discontinuity of the gauge field is due to the sudden kick of particles. If we consider smooth trajectories, the obtained potential is also smooth. Our justification to consider the trajectory \eqref{initial_tr} and \eqref{final_tr} relies on the leading and the subleading soft theorems in QED. These soft theorems state that the soft factors are determined by the initial momenta and angular momenta of particles and do not depend on the details of the scattering. Thus, in order to see the memory effect associated with the soft theorems, only the initial and final (angular) momenta of particles are relevant. Thus we expect that the details of smearing do not affect the memory effects. 

We expand this potential in the large $r$ limit  with $u=t-r$ fixed.\footnote{$r=|\vec{x}|$. $\hat{\vec{x}}=\vec{x}/r$.} 
The potential is expanded as 
\begin{align}
A^\mu(x)=&\Theta(u+\vec{x}_0\cdot \hat{\vec{x}}-t_0 )\sum_{n'} \frac{e_{n'} p_{n'}^\mu}{4\pi \omega_{n'} r (1-\vec{v}_{n'}\cdot \hat{\vec{x}})}
\nn
&+\Theta(-u-\vec{x}_0\cdot \hat{\vec{x}}+t_0 )\sum_{n} \frac{e_{n} p_{n}^\mu}{4\pi \omega_{n} r (1-\vec{v}_{n}\cdot \hat{\vec{x}})}
+\mathcal{O}(r^{-2})
\nn
=&\Theta(u+\vec{x}_0\cdot \hat{\vec{x}}-t_0 )\sum_{n'} \frac{e_{n'} p_{n'}^\mu}{4\pi p_{n'}^u r}
+\Theta(-u-\vec{x}_0\cdot \hat{\vec{x}}+t_0 )\sum_{n} \frac{e_{n} p_{n}^\mu}{4\pi p_{n}^u r}
+\mathcal{O}(r^{-2}).
\label{eq:em_cl_sol}
\end{align}

\red{The leading electromagnetic memory is the shift of the gauge field from $u=-\infty$ to $u=\infty$  at large $r$ region (on $\mathscr{I}^+$): 
	\begin{align}
	 A^\mu(u=+\infty) -A^\mu(u=-\infty) =\int^\infty_{-\infty} du \partial_u A^\mu \,.
	\end{align}
	For our case \eqref{eq:em_cl_sol}, it is given by 
	\begin{align}\label{Eq:U(1)_leading_memory}
	\int^\infty_{-\infty} du \partial_u A^\mu  =\sum_{n'} \frac{e_{n'} p_{n'}^\mu}{4\pi p_{n'}^u r} -\sum_{n} \frac{e_{n} p_{n}^\mu}{4\pi p_{n}^u r}\,.
	\end{align}
	The right hand side is certainly the soft factor in the leading soft photon theorem. 
	
	The subleading memory is related to the integral $\int^\infty_{-\infty} du\, u \partial_u A^\mu$ on $\mathscr{I}^+$ \cite{Lysov:2014csa, Campiglia:2016hvg, Conde:2016csj}.
	Using eq.~\eqref{eq:em_cl_sol}, the integral is computed as 
	\begin{align}
	\int^\infty_{-\infty} du\, u \partial_u A^\mu =(t_0-\vec{x}_0\cdot \hat{\vec{x}})\left[
	\sum_{n'} \frac{e_{n'} p_{n'}^\mu}{4\pi p_{n'}^u r}
	-\sum_{n} \frac{e_{n} p_{n}^\mu}{4\pi p_{n}^u r} 
	\right] \,.
	\label{em_sub_int}
	\end{align}
	Since the angular momenta of particles with trajectories \eqref{initial_tr} and \eqref{final_tr} are given by\footnote{We have the ambiguity of the choice of the origin to define angular momenta. In fact, angular momenta are not invariant under spacetime translations. It is related to the ambiguity to define the subleading memory as  
		$\int^\infty_{-\infty} du\, (u-u_0) \partial_u A^\mu$ where $u_0$ is arbitrary. 
		In the absence of the magnetic charged object, by concentrating the $B$-mode part, this ambiguity is removed.}
	\begin{align}
	J_n^{\mu\nu}=x_0^\mu p_n^\nu-x_0^\nu p_n^\mu\,, \quad J_{n'}^{\mu\nu}=x_0^\mu p_{n'}^\nu-x_0^\nu p_{n'}^\mu\,,
	\end{align}
	we have 
	\begin{align}
	J_n^{u \mu}=(t_0-\vec{x}_0\cdot \hat{\vec{x}})p_n^\mu-p_n^u x_0^\mu\,, \quad  
	J_{n'}^{u \mu}=(t_0-\vec{x}_0\cdot \hat{\vec{x}})p_{n'}^\mu-p_{n'}^u x_0^\mu \,. 
	\end{align}
	Eq.~\eqref{em_sub_int} is then rewritten as 
	\begin{align}
	&\left[
	\sum_{n'} \frac{e_{n'} J_{n'}^{u\mu}}{4\pi p_{n'}^u r}
	-\sum_{n} \frac{e_{n} J_{n}^{u\mu}}{4\pi p_{n}^u r}\right]
	+\left[\sum_{n'} e_{n'}
	-\sum_{n} e_{n} \right]\frac{x_0^\mu}{4\pi r}
	\nn
	&
	=\left[
	\sum_{n'} \frac{e_{n'} J_{n'}^{u\mu}}{4\pi p_{n'}^u r}
	-\sum_{n} \frac{e_{n} J_{n}^{u\mu}}{4\pi p_{n}^u r}\right]\,,
	\end{align}
	where we have used the charge conservation.
	Thus, the subleading memory is given by 
		\begin{align}\label{Eq:U(1)_subleading_memory}
	\int^\infty_{-\infty} du\, u \partial_u A^\mu =
	\sum_{n'} \frac{e_{n'} J_{n'}^{u\mu}}{4\pi p_{n'}^u r}
	-\sum_{n} \frac{e_{n} J_{n}^{u\mu}}{4\pi p_{n}^u r} \,,
	\end{align}
	which is the same form as the soft factor in the subleading soft theorem  
	as discussed in the earlier works~\cite{Mao:2017wvx}. 
}

\red{
We can see that these leading and subleading memories appears in the classical radiation as characteristic $u$-dependences. 
In fact, if we further expand \eqref{eq:em_cl_sol}} with respect to $\vec{x}_0\cdot \hat{\vec{x}}-t_0$,\footnote{\red{As far as we know, this view point was not mentioned before.} See Refs.~\cite{Caboz,Estrada} for the discussions of the Taylor expansions for distributions from a mathematical perspective.}  it is written as
\begin{align}
\label{eq:em_rad_memory}
A^\mu(x)=&\Theta(u)\sum_{n'} \frac{e_{n'} p_{n'}^\mu}{4\pi p_{n'}^u r}
+\Theta(-u)\sum_{n} \frac{e_{n} p_{n}^\mu}{4\pi p_{n}^u r}
-\delta(u)\left[
\sum_{n'} \frac{e_{n'} J_{n'}^{u\mu}}{4\pi p_{n'}^u r}
-\sum_{n} \frac{e_{n} J_{n}^{u\mu}}{4\pi p_{n}^u r}\right]
\nn
&+\frac{1}{r}\mathcal{O}((\vec{x}_0\cdot \hat{\vec{x}}-t_0 )^2) +\mathcal{O}(r^{-2}).
\end{align}
\red{
Therefore, for the radiation by the kick \eqref{Eq:elemag_source}, the leading and subleading memories are related to the coefficient of the step function $\Theta(u)$ and  the delta function $\delta(u)$, respectively.\footnote{
	\red{This expansion can be used for the case that $\abs{u}$ is large. The expression \eqref{eq:em_rad_memory} means that, when we characterize the long time behavior of the radiation w.r.t.~$u$, the first approximation is the step function which is related to the leading memory, and next correction is related to the subleading memory.}
}
We will use this fact in Sec.~\ref{Sec:detection_memory}.
}

\subsubsection{$E$ and $B$ mode decomposition}
The decomposition of the field to parity odd and even modes is sometimes useful. The classical example may be the observation of the fluctuation of the cosmic microwave background. From the detection of the $E$-mode, we learned the information of the density perturbation. The future observation of the $B$-mode may provide us the information of the tensor perturbation.
Similarly, here we consider the $E, B$ decomposition of the memory effect.
The definition of the $E,B$ mode decomposition is reviewed in App.~\ref{App:EB_decomposition}. 

Since $p_B/(r p^u)= -\partial_B \log(1-\vec{v}\cdot\hat{\vec{x}})$, 
the spherical components of Eq.~\eqref{Eq:U(1)_leading_memory} corresponds to the $E$-mode,
\al{\label{Eq:leading_E_memory}
&A_B(u=\infty)-A_B(u=-\infty)\nn
&= -\frac{1}{4\pi} \p_B \left[\sum_{n'} e_{n'}\log\paren{1-\vec{v}_{n'}\cdot\hat{\vec{x}}}
-\sum_{n} e_{n} \log\paren{1-\vec{v}_{n}\cdot\hat{\vec{x}}}\right]+\mathcal{O}\paren{1\over r}.
}
Thus, the leading memory is the $E$-mode, which is consistent with the Ref.~\cite{Winicour:2014ska}. In Ref.~\cite{Winicour:2014ska}, it is claimed that the physically reasonable source only gives 
the E-mode leading memory. However, this is too strong statement. If we allow the scattering of magnetic monopoles, this conclusion is modified. The soft theorem in the presence of the magnetic charge is investigated in Refs.~\cite{Strominger:2015bla,Strominger:2017zoo}. 
For a positive-helicity soft photon, the soft factor becomes
\al{
\sum_k {p_k\cdot\paren{e_k\epsilon_z+g_k \tilde{\epsilon}_z}\over p_k\cdot q},
}
where $e_k$ and $g_k$ are the electric and magnetic charge, respectively, and $\epsilon, \tilde{\epsilon}$ are the polarization vectors of the original and dual $U(1)$ gauge fields.
In the neighborhood of the future null infinity $\mathcal{I}^+$, 
we can see that $\tilde{\epsilon}_z=i \epsilon_z$~\cite{Strominger:2015bla}.
Combining with Eq.~\eqref{Eq:leading_E_memory} and the fact that the memory effect is equivalent to the soft photon theorem, it can be seen that the leading B-mode memory is generated by the magnetic monopole.  

Next, let us consider the subleading memory effect, Eq.~\eqref{Eq:U(1)_subleading_memory}.
\al{\label{Eq:FuB_laeding}
F_{uB}&\simeq
-{1\over r}\delta'(u){J^u_B\over\omega\paren{1-\vec{v}\cdot\hat{x}}}=-{1\over r}\delta'(u){b^u p_B-b_B p^u\over\omega\paren{1-\vec{v}\cdot\hat{x}}}
\nn
&=-\delta'(u)\br{
{b^t-\vec{b}\cdot\hat{x}\over\omega\paren{1-\vec{v}\cdot\hat{x}}}\paren{\p_B\hat{x}}\cdot\vec{p}-\paren{\p_B\hat{x}}\cdot\vec{b}
}
.
}
Here, we write
\al{
J_{\mu\nu}=b_\mu p_\nu-b_\nu p_\mu.
}
Without loss of generality, $b_\mu$ does not contain the component 
proportional to $p_\mu$.
We calculate $\p_\theta F_{u\varphi}-\p_\varphi F_{u\theta}$ to extract the B-mode contribution:
\al{
\p_\theta F_{u\varphi}-\p_\varphi F_{u\theta}=\delta'(u){1\over 1-\vec{v}\cdot\hat{x}}\paren{\p_\theta\hat{x}^i}\paren{\p_\varphi\hat{x}^j}\paren{b_iv_j-b_j v_i}\neq0.
}
Therefore, the subleading memory contains the B-mode. Similarly, if the external particle is the magnetic monopole, the subleading memory contains the $E$-mode.

\subsection{Gravity}
\subsubsection{Derivation of the gravitational memory effect}
We repeat a similar analysis for the gravitational potential.  
We consider linearized gravitational potential $h_{\mu\nu}$ from the above motion of particles \eqref{initial_tr} and \eqref{final_tr}. The stress-energy tensor is given by
\begin{align}\label{Eq:gravity_source}
T^{\mu\nu}(x)= \sum_n \int^0_{-\infty} d \tau\, \frac{p_n^\mu p_n^\nu}{m_n}\, \delta^{(4)}(x-y_n(\tau))
+\sum_{n'} \int^{\infty}_0 d \tau\, \frac{p_{n'}^\mu p_{n'}^\nu}{m_{n'}}\, \delta^{(4)}(x-y_{n'}(\tau)).
\end{align}
In the harmonic gauge (the de Donder gauge)\footnote{Explicitly, the gauge condition is $\p_\mu h^{\mu\nu}-\frac12 \p^\nu h^\mu_\mu=0$. }, the linearized Einstein equation takes the form  
\begin{align}
\Box \bar{h}^{\mu\nu}= -16 \pi G_N T^{\mu\nu}\,, \quad \text{where}\quad \bar{h}^{\mu\nu}= h^{\mu\nu}-\frac12 \eta^{\mu\nu} h^{\lambda}_{\lambda}\,. 
\end{align}
The retarded solution is given by\footnote{The original gravitational potential $h^{\mu\nu}$ is obtained by 
	$h^{\mu\nu}= \bar{h}^{\mu\nu}-\frac12 \eta^{\mu\nu} \bar{h}^{\lambda}_{\lambda}$.} 
\begin{align}
&\frac{1}{16 \pi G_N}\bar{h}^{\mu\nu}(x)\nn
&=\Theta(|\vec{x}-\vec{x}_0|-t+t_0)\sum_n \frac{p_n^\mu p_n^\nu}{4\pi m_n\sqrt{|\vec{x}-\vec{x}_0|^2-\paren{\frac{\vec{v}_n\cdot(\vec{x}-\vec{x}_0)}{v_n}}^2+\frac{1}{1-v_n^2}\paren{\frac{\vec{v}_n\cdot(\vec{x}-\vec{x}_0)}{v_n}-v_n (t-t_0)}^2}}
\nn
&+\Theta(-|\vec{x}-\vec{x}_0|+t-t_0)\sum_{n'} \frac{p_{n'}^\mu p_{n'}^\nu}{4\pi m_{n'}\sqrt{|\vec{x}-\vec{x}_0|^2-\paren{\frac{\vec{v}_{n'}\cdot(\vec{x}-\vec{x}_0)}{v_{n'}}}^2+\frac{1}{1-v_{n'}^2}\paren{\frac{\vec{v}_{n'}\cdot(\vec{x}-\vec{x}_0)}{v_{n'}}-v_{n'} (t-t_0)}^2}}
\,.
\end{align}
The original gravitational potential $h^{\mu\nu}$ is obtained by 
\begin{align}
h^{\mu\nu}= \bar{h}^{\mu\nu}-\frac12 \eta^{\mu\nu} \bar{h}^{\lambda}_{\lambda}.
\end{align}
In the large $r$ limit with $u=t-r$ fixed, it is expanded as 
\begin{align}
\frac{1}{16 \pi G_N}\bar{h}^{\mu\nu}(x)=&\Theta(u+\vec{x}_0\cdot \hat{\vec{x}}-t_0 )\sum_{n'} \frac{p_{n'}^\mu p_{n'}^\nu}{4\pi p_{n'}^u r}
+\Theta(-u-\vec{x}_0\cdot \hat{\vec{x}}+t_0 )\sum_{n} \frac{p_n^\mu p_n^\nu}{4\pi p_{n}^u r}
+\mathcal{O}(r^{-2}).
\end{align} 

If we further expand it with respect to $\vec{x}_0\cdot \hat{\vec{x}}-t_0$, it is written as 
\begin{align}
\label{gw_soft}
\frac{1}{16 \pi G_N}\bar{h}^{\mu\nu}(x)=&\Theta(u)\sum_{n'} \frac{p_{n'}^\mu p_{n'}^\nu}{4\pi p_{n'}^u r}
+\Theta(-u)\sum_{n} \frac{p_n^\mu p_n^\nu}{4\pi p_{n}^u r}
+\delta(u)(\vec{x}_0\cdot \hat{\vec{x}}-t_0 )\left[
\sum_{n'} \frac{p_{n'}^\mu p_{n'}^\nu}{4\pi p_{n'}^u r}
-\sum_{n} \frac{p_n^\mu p_n^\nu}{4\pi p_{n}^u r}
\right]
\nn
&+\frac12 \delta'(u)(\vec{x}_0\cdot \hat{\vec{x}}-t_0 )^2\left[
\sum_{n'} \frac{p_{n'}^\mu p_{n'}^\nu}{4\pi p_{n'}^u r}
-\sum_{n} \frac{p_n^\mu p_n^\nu}{4\pi p_{n}^u r}
\right]
+\frac{1}{r}\mathcal{O}\paren{(\vec{x}_0\cdot \hat{\vec{x}}-t_0 )^3} +\mathcal{O}(r^{-2}).
\end{align}
Using the momentum conservation, the third term of eq.\eqref{gw_soft} can be rewritten as follows: 
\begin{align}
&-\delta(u)\left[
\sum_{n'} \frac{p_{n'}^\nu J_{n'}^{u\mu}}{4\pi p_{n'}^u r}
-\sum_{n} \frac{p_{n}^\nu J_{n}^{u\mu}}{4\pi p_{n}^u r}\right]
-\delta(u)\left[\sum_{n'} p_{n'}^\nu
-\sum_{n} p_{n}^\nu \right]\frac{x_0^\mu}{4\pi r}
=-\delta(u)\left[
\sum_{n'} \frac{p_{n'}^\nu J_{n'}^{u\mu}}{4\pi p_{n'}^u r}
-\sum_{n} \frac{p_{n}^\nu J_{n}^{u\mu}}{4\pi p_{n}^u r}\right]. 
\end{align}
The second line of eq.\eqref{gw_soft} can be written as 
\begin{align}
&\frac12 \delta'(u)\left[
\sum_{n'} \frac{J_{n'}^{u\mu}J_{n'}^{u\nu}}{4\pi p_{n'}^u r}
-\sum_{n} \frac{J_{n}^{u\mu}J_{n}^{u\nu}}{4\pi p_{n}^u r}\right]
+\frac12 \delta'(u)\left[\sum_{n'}J_{n'}^{u\nu}
-\sum_{n} J_{n}^{u\nu}\right]\frac{x_0^\mu}{4\pi r}
\nn
&+\frac12 \delta'(u)\left[\sum_{n'}J_{n'}^{u\mu}
-\sum_{n} J_{n}^{u\mu}\right]\frac{x_0^\nu}{4\pi r}
+\frac12 \delta'(u)\left[\sum_{n'}p_{n'}^{u}
-\sum_{n} p_{n}^{u}\right]\frac{x_0^\mu x_0^\nu}{4\pi r}
\nn
&=\frac12 \delta'(u)\left[
\sum_{n'} \frac{J_{n'}^{u\mu}J_{n'}^{u\nu}}{4\pi p_{n'}^u r}
-\sum_{n} \frac{J_{n}^{u\mu}J_{n}^{u\nu}}{4\pi p_{n}^u r}\right]\,, 
\end{align}
where we have used the momentum conservation and the angular momentum conservation.
Thus, we have 
\begin{align}
\frac{1}{16 \pi G_N}\bar{h}^{\mu\nu}(x)=&\Theta(u)\sum_{n'} \frac{p_{n'}^\mu p_{n'}^\nu}{4\pi p_{n'}^u r}
+\Theta(-u)\sum_{n} \frac{p_n^\mu p_n^\nu}{4\pi p_{n}^u r}
-\delta(u)\left[
\sum_{n'} \frac{p_{n'}^\nu J_{n'}^{u\mu}}{4\pi p_{n'}^u r}
-\sum_{n} \frac{p_{n}^\nu J_{n}^{u\mu}}{4\pi p_{n}^u r}\right]
\nn
&+\frac12 \delta'(u)\left[
\sum_{n'} \frac{J_{n'}^{u\mu}J_{n'}^{u\nu}}{4\pi p_{n'}^u r}
-\sum_{n} \frac{J_{n}^{u\mu}J_{n}^{u\nu}}{4\pi p_{n}^u r}\right]
+\frac{1}{r}\mathcal{O}\paren{(\vec{x}_0\cdot \hat{\vec{x}}-t_0 )^3} +\mathcal{O}(r^{-2}).
\label{230}
\end{align}
The step function $\Theta(\pm u)$, the delta function $\delta(u)$ and $\delta'(u)$ terms correspond to the leading, subleading and subsubleading memory effects, respectively. These memory effects are characterized by the following integrations:
\al{\label{Eq:gravity_memory_integrals}
&\int^\infty_{-\infty}du \p_u h_{\mu\nu},
&&\int^\infty_{-\infty}du u \p_u h_{\mu\nu},
&&\int^\infty_{-\infty}du u^2 \p_u h_{\mu\nu}.
}
We note that, as in the electromagnetic case, we have an ambiguity to choose
\al{
&\int^\infty_{-\infty}du \paren{u-u_0} \p_u h_{\mu\nu},
&&\int^\infty_{-\infty}du \paren{u-u_0}^2 \p_u h_{\mu\nu}.
}
Regarding the subleading order, this ambiguity is removed if one focuses on the $B$-mode part, while the ambiguity of the subsubleading order remains.

\subsubsection{$E,B$ decomposition}
Let us consider the $E,B$ decomposition of the memory effect.
The leading gravitational memory effect is always the $E$-mode because the following relation holds:
\al{
{p_z^2\over 1-\vec{v}\cdot\hat{x}}=\omega^2 r^2 D_z^2\br{\paren{1-\vec{v}\cdot\hat{x}}\ln\paren{1-\vec{v}\cdot\hat{x}}}.
}
Note that the relation,
\al{
D_z^2 \hat{x}=\paren{\p_z+{2\bar{z}\over1+|z|^2}}\p_z\hat{x}=0,
}
has been used.
Contrary to the $U(1)$ cases, there are no sources which lead to the leading B-mode memory, because every object universally couples with gravity. Therefore, the leading gravitational memory is always the $E$-mode memory.

As in the $U(1)$ case, the subleading part contains the $B$-mode memory. The nonzero $D_{\bar{z}}^2 h_{zz}-D_z^2 h_{\bar{z}\bar{z}}$ means the existence of the $B$-mode (see App.~\ref{App:EB_decomposition}), and we can see
\al{
D_{\bar{z}}^2{p_z J^u_z\over 1-\vec{v}\cdot\hat{x}}-D_z^2{p_{\bar{z}} J^u_{\bar{z}}\over 1-\vec{v}\cdot\hat{x}}&=
\omega(D_{\bar{z}}^2 p_zb_z-D_z^2p_{\bar{z}}b_{\bar{z}})+\br{D_z^2\paren{{1-\vec{b}\cdot\hat{x}\over 1-\vec{v}\cdot\hat{x}}p_{\bar{z}}^2}-D_{\bar{z}}^2\paren{{1-\vec{b}\cdot\hat{x}\over 1-\vec{v}\cdot\hat{x}}p_z^2}}
\nn&=D_z^2\paren{{1-\vec{b}\cdot\hat{x}\over 1-\vec{v}\cdot\hat{x}}p_{\bar{z}}^2}-D_{\bar{z}}^2\paren{{1-\vec{b}\cdot\hat{x}\over 1-\vec{v}\cdot\hat{x}}p_z^2}.
}
In general, this is not zero, and therefore the subleading memory contains the B-mode. 

\subsection{Axion}
\subsubsection{Derivation of the axion memory effect}
The radiated part of axion field $a$ corresponding to the subleading soft theorem takes the following form in the momentum space: 
\al{\label{Eq:axion_solution}
\tilde{a}(q)={1\over|\vec{q}|}\delta(q^0-|\vec{q}|)\epsilon_{\mu\nu\rho\sigma}\paren{{q^\mu p'^\nu  J_{p'}^{\rho\sigma}\over p'\cdot q}-{q^\mu p^\nu  J_{p}^{\rho\sigma}\over p\cdot q}},
}
which can be derived by using the equivalence between the soft theorem and the classical field. In Ref.~\cite{Hamada:2017atr}, the subleading soft theorem corresponding to Eq.~\eqref{Eq:axion_solution} is proved for spin $0$ and $1/2$ external particles. In App.~\ref{App:soft_pion_theorem_spin1}, we present the proof for spin $1$ particle.
Performing the Fourier transformation of \eqref{Eq:axion_solution}, we obtain
\al{\label{Eq:axion memory}
a(x)&=
\epsilon_{\mu\nu\rho\sigma}{1\over r}\p^\mu\paren{\Theta(u){p'^\nu J_{p'}^{\rho\sigma}\over \omega_{p'}\paren{1-\vec{v}'\cdot\hat{x}}}-\Theta(-u){p^\nu J_{p}^{\rho\sigma}\over \omega_{p}\paren{1-\vec{v}\cdot\hat{x}}}}+\mathcal{O}(r^{-2})
\nn&=-\epsilon^u_{\nu\rho\sigma}{1\over r}\delta(u)\paren{{p'^\nu J_{p'}^{\rho\sigma}\over p'_r}+{p^\nu J_{p}^{\rho\sigma}\over p_r}}+\mathcal{O}(r^{-2})
\nn&=: {1\over r}\delta(u) S^{(1)}+{1\over r^2}\Theta(u) S^{(2)}.
}
Therefore, the coefficient of $1/r$ is zero at past and future.\footnote{\red{In Ref.~\cite{Hamada:2017atr}, we discussed the memory effect of $r^{-2}\Theta(u)$ term.}} 
We note that, since the axion field is scalar, there is no notion of the $E, B$ mode decomposition. 


\section{Detecting the memory effects}\label{Sec:detection_memory}
So far, we have seen that the memory effects are encoded in $\Theta(u), \delta(u)$ and $\delta'(u)$ terms in the radiation field \red{when we see the long time behavior}. Here we discuss how to detect the memory effects. It is known that the memory effect can be probed by looking the change of the trajectory of the free falling particle. In addition to this, we show that the direction of the spin can be a good observable to detect the memory effects.

\subsection{Electromagnetic memory}
\subsubsection{The memory of the trajectory}
The non-relativistic equation of motion of a charged test particle is
\al{
M\ddot{\vec{x}}=Q\left( \vec{E}+\dot{\vec{x}}\times\vec{B}\right),
}
where the dot means the derivative with respect to $t$, $Q$ is the charge of the test particle, and $\vec{E}_i=F_{i0}$ and $\vec{B}_i={1\over2}\epsilon_{ijk}F^{jk}$ are given by
\al{\label{Eq:EB_field}
&\vec{E}=\delta(u)\br{-\hat{r}_i \paren{S^{(0A)}_0-S^{(0B)}_0}+\paren{S^{(0A)}_i-S^{(0B)}_i}}-\delta'(u)\paren{\hat{r}_iS_0^{(1)}+S^{(1)}_i}+\mathcal{O}\paren{1\over r^2},
\nn&
\vec{B}={1\over2}\epsilon_{ijk}
\sqbr{\delta(u)\br{-\paren{\hat{r}^iS^{(0A)j}-\hat{r}^jS^{(0A)i}}+\paren{\hat{r}^iS^{(0B)j}-\hat{r}^jS^{(0B)i}}}-\delta'(u)\paren{\hat{r}^i S^{(1)j}-\hat{r}^j S^{(1)i}}}
\nn&\phantom{\vec{B}=}+\mathcal{O}\paren{1\over r^2}, 
}
where we have used $A_\mu\simeq\Theta(u)S^{(0A)}_\mu+\Theta(-u)S^{(0B)}_\mu+\delta(u)S^{(1)}_\mu$.
Then, we can see
\al{
M\ddot{x}_k=C_k\delta(u)+D_k\delta'(u),
}
where
\al{&
C_k=-Q\left[\br{\hat{r}_k \paren{S^{(0A)}_0-S^{(0B)}_0}+\paren{S^{(0A)}_k-S^{(0B)}_k}}
+\dot{x}_i\paren{-\hat{r}^{[i}S^{(0A)k]}+\hat{r}^{[i}S^{(0B)k]}}\right]
,
\nn&
D_k=-Q\left[\paren{\hat{r}_iS_0^{(1)}+S^{(1)}_i}-\dot{x}_i\hat{r}^{[i}S^{(1B)k]}\right].
}
Here $D^{[ij]}$ means ${1\over2}\paren{D^{ij}-D^{ji}}$. 
The velocity of the test particle is constant for $u<0$ and $u>0$.
The jumps of the velocity and the position at $u=0$ can be calculated as
\al{\label{Eq:elemag_memory_detection}&
\Delta \dot{x}_k= {C_k\over M}\Bigg|_{(r,\theta,\varphi)=(r_*,\theta_*,\varphi_*)},
&&
\Delta x_k= {D_k\over M}\Bigg|_{(r,\theta,\varphi)=(r_*,\theta_*,\varphi_*)},
}
where $(r_*,\theta_*,\varphi_*)$ are the position of the test particle at $u=0$,\footnote{We assume that $\Delta x_k$ is much smaller than $r_*$.} and $\Delta \dot{x}_k:=\lim_{\epsilon\to0}\paren{\dot{x}_k|_{t=r_*+\epsilon}-\dot{x}_k|_{t=r_*-\epsilon}}$, $\Delta x_k:=\lim_{\epsilon\to0}\paren{x_k|_{t=r_*+\epsilon}-x_k|_{t=r_*-\epsilon}}$ .
In this way, the leading memory effect can be detected as a kick memory of the test particles~\cite{Bieri:2013hqa}. The subleading memory effect as the jump of the coordinate $x$ was pointed out in Ref.~\cite{Mao:2017axa}. 
Although it seems difficult to detect the subleading memory effect, the leading memory drops out if we take a particular combination by multiplying the projection operator, \textit{i.e.}, only the $B$-mode contribution can be picked up, see Ref.~\cite{Mao:2017axa} for the detail.

\subsubsection{The memory of the spin}
If the test particle has a spin, it is natural to consider the time evolution of the spin due to the pulse injection. For simplicity, we consider a non-relativistic particle. Then, the interaction Hamiltonian is 
\al{
H_\text{int}=\vec{B}\cdot\vec{s},
}
where $\vec{s}$ is the spin operator. The Heisenberg equation of the spin operator is
\al{&
\dot{\vec{s}}=\vec{B}\times\vec{s},
\quad \to
\quad \vec{s}=\int dt\paren{\vec{B}\times\vec{s}}.
}
The magnetic field $\vec{B}$ would be written as $\vec{B}={\vec{B_1}\over r}\delta(u)+{\vec{B_2}\over r}\delta'(u)$ (see Eq.~\eqref{Eq:EB_field}). Then, the above equation means that the spin rotation happens by the leading effect, $B_1$. Thus, the leading memory is related to the change of the direction of the spin. The subleading part does not contribute to such a permanent change of the spin at $\mathcal{O}(1/r)$ order.

\subsection{Gravitational memory}
\subsubsection{The memory of the trajectory and time delay}
Here we briefly review the detection of the leading and subleading memory effect by measuring the trajectory of the particle and the time delay.
One can easily see that the leading memory effect can be probed by the displacement memory effect between two free particles~\cite{ZelPol,Braginsky:1986ia,Braginsky1987,Christodoulou:1991cr}. The proper distance at $u>0$ is different from the distance at $u<0$. It is non-trivial to detect the subleading memory effect, namely the $\delta(u)$ part. However, in Ref.~\cite{Pasterski:2015tva}, it is shown that the Sagnac effect can be used to probe the subleading memory effect. The effect from the leading memory can be removed because only the B-mode part contributes to the Sagnac effect.

Then, the natural question is whether we can detect the subsubleading memory effect by only looking the observable at far past and future. As shown in Sec.~\ref{Sec:memory_effect}, formally, the subsubleading memory is characterized by
\al{\label{eq:susbsub_int}
\int^\infty_{-\infty}du u^2 \p_u h_{\mu\nu},
}
which corresponds to the coefficient of $\delta'(u)$ \red{in the classical gravitational wave produced by the kick of particles \eqref{Eq:gravity_source}.
One might think that the finiteness of the integral \eqref{eq:susbsub_int} is owing to our specific choice of the source. 
However, for the same reason below eq.~\eqref{eq:em_ret_sol}, we expect that the integral is also finite for other trajectories of particles as long as the duration time of the burst is finite, and the value is characterized only by the momenta and angular momenta.} 

\red{
Unfortunately, as far as we investigate, for a massive test particle, the subsubleading gravitational memory effect cannot be observed by measuring the trajectories at far future and past. 
We consider that  the massive particle is first at rest with large $r$ in the background $\eta_{\mu\nu}+h_{\mu\nu}$, where $h_{\mu\nu}$ has a form Eq.~\eqref{230}. Due to this form, the velocity of the particle after the burst is order $r^{-1}$. 
The geodesic equation with respect to the proper time $\tau$, 
\begin{align}
\frac{d^2 x^\mu}{d\tau^2} + \Gamma^\mu_{\phantom{\mu}\nu\rho} \frac{d x^\nu}{d\tau} \frac{d x^\rho}{d\tau}=0\,,
\end{align}
is approximated as 
\begin{align}
\frac{d^2 x^i}{d\tau^2} = -\Gamma^i_{\phantom{i}tt} \frac{d t}{d\tau} \frac{d t}{d\tau} ,
\label{free_eom}
\end{align}
by taking up to the $\mathcal{O}(r^{-1})$ terms. Here, $\Gamma^i_{\phantom{i}tt}$ takes the form 
\begin{align}
\Gamma^i_{\phantom{i}tt} = \delta(u)\Delta^{i(0)}+\delta'(u)\Delta^{i(1)}+\delta''(u)\Delta^{i(2)}+ \mathcal{O}(r^{-2}),
\end{align}
where $\Delta^{i(0)}, \Delta^{i(1)}$ and $\Delta^{i(2)}$ are the factors related to the leading, subleading and subsubleading soft factors, respectively.
By integrating the EoM \eqref{free_eom}, we can see that $\Delta^{i(0)}$ is related to the shift of the velocity (the kick memory) and $\Delta^{i(1)}$ corresponds to the position memory. However, the subsubleading term $\Delta^{i(2)}$ does not lead to such a step function. Thus, the effect from $\Delta^{i(2)}$ cannot be detected as a memory for the trajectory of the test particle. 
}

\subsubsection{The memory of the spin}
As in the electromagnetic memory, we consider the non-relativistic Hamiltonian of the spin $1/2$ particle. In order to derive the non-relativistic Hamiltonian including the interaction with the linearized gravitational field, we first consider the Dirac equation given by\footnote{See also Ref.~\cite{Deriglazov:2017jub} for the equations of motion of the spin in the curved background.} 
\al{
({\Slash \nabla}+m)\psi=0,
}
where 
\begin{align}
&{\Slash \nabla}\psi = {\Slash \p} \psi +\frac12 h^{\mu\nu}\gamma_\nu \p_\mu \psi+\frac{1}{4}(\p_\mu h_{\nu\rho})(\eta^{\rho\mu}\gamma^\nu+\eta^{\rho\nu}\gamma^\mu)\psi\,,
&&
	\gamma^0=
	\begin{pmatrix}
		-i&0\\
		0&i\\
	\end{pmatrix},
	\quad
	\gamma^i=
	\begin{pmatrix}
		0&-i\sigma^i\\
		i\sigma^i&0\\
	\end{pmatrix}.
\end{align}
In this subsection, we concentrate on the configuration where $h_{\mu\nu}$ satisfies the transverse traceless condition, that is,
\al{
h_{0\mu}=h_{ii}=\p_i h_{ij}=0.
}
Then, the Dirac equation reduces to 
\al{
\paren{{\Slash \p}+{1\over2} h^{ij}\gamma_j \p_i +m}\psi=0.
}
By employing the following parameterization,
\al{
\psi=e^{-imt}
\begin{pmatrix}
\varphi\\ \xi
\end{pmatrix},
}
we perform the non-relativistic approximation.
In this parametrization, the Dirac equation becomes
\begin{align}
\gamma^0\p_0 
\begin{pmatrix}
\varphi\\ \xi
\end{pmatrix}
= \left\{ im \left[(i+\gamma^0) \right]-\gamma^i \p_i -\frac12 h^{ij}\gamma_j \p_i
\right\}
\begin{pmatrix}
\varphi\\ \xi
\end{pmatrix}. 
\end{align}
In the non-relativistic limit, we get
\al{
\xi\simeq -{i\over 2m}\sigma^i \p_i \varphi - {i\over4m}h^{ij}\sigma_i\p_j\varphi,
}
from which we arrive at
\al{
i\p_0\varphi=-{1\over2m}\p_i^2\varphi-{1\over2m}h^{ij}\p_i\p_j\varphi-{1\over8m}[\sigma^i,\sigma_j](\p_i h^{jk})\p_k\varphi.
}
The right hand side can be interpreted as the non-relativistic Hamiltonian. The last term corresponds to the interaction of the spin and gravity. 
Since the spin operator does not commute with the last term, in general, the direction of the spin changes by the pulse emission. More concretely, the Heisenberg equation of the spin operator $s^n=\sigma^n/2$ is
\al{
\dot{s}^n=-{1\over2m}\paren{\delta^{il}\delta^{jn}-\delta^{in}\delta^{jl}}s^l(\p_i h_j^k)p_k.
}
The derivative of $h_{ij}$ contains the delta function at order $r^{-1}$, and by integrating over time, we would get the permanent change of the direction of the spin. One can see that this spin change can detect the leading memory effect which is proportional to the step function, and cannot detect the subleading and subsubleading memory effects.

\subsection{Axion memory}
Here we consider the the detection of the axion memory effect. We take spin $1/2$ or $1$ particle as a probe.
\subsubsection{Coupling with fermion}
The non-relativistic Hamiltonian for the fermion which couples with the axion $a$ is
\al{
H_\text{NR}=
-{p^2\over2M}
-\kappa (\vec{\sigma}\cdot\vec{\nabla} a).
}
Therefore, the spin couples with the axion field, and it is expected that the axion memory effect can be proved by observing the change of the direction of spin.
The time evolutions of the momentum and spin are given by Heisenberg equation:
\al{&
\dot{\vec{p}}={1\over2}\kappa \vec{\nabla}(\vec{s}\cdot\vec{\nabla} a),
&&
\dot{\vec{s}}=-{1\over2}\kappa(\vec{\nabla}\times\vec{s}) a.
}
We can solve it easily, and obtain
\al{&
\vec{p}(\infty)-\vec{p}(-\infty)={1\over2}\kappa\int dt \br{ \vec{\nabla}(\vec{s}\cdot\vec{\nabla}) a},
&&
\vec{s}(\infty)-\vec{s}(-\infty)=-{1\over2}\kappa \int dt \br{(\vec{\nabla}\times \vec{s}) a}.
}
Hence, the change of the spin happen at the order of $r^{-2}$. 
Explicitly, by putting Eq.~\eqref{Eq:axion memory}, we get 
\al{
&\vec{p}(\infty)-\vec{p}(-\infty)={\kappa\over2}\sqbr{\vec{\nabla}(\vec{s}\cdot\vec{\nabla})\paren{{S^{(1)}\over r}}-\vec{\nabla}\br{\paren{\vec{s}\cdot\hat{r}}{S^{(2)}\over r^2}}-\hat{r}\paren{\vec{s}\cdot\vec{\nabla}}\paren{{S^{(2)}\over r^2}}}\Bigg|_{(r_*,\theta_*,\varphi_*)}+\mathcal{O}\paren{1\over r^4},
\nn&
\vec{s}(\infty)-\vec{s}(-\infty)={\kappa\over2}\sqbr{-\paren{\vec{\nabla}\times\vec{s}}\paren{S^{(1)}\over r}+\paren{\hat{r}\times\vec{s}}{S^{(2)}\over r^2}}\Bigg|_{(r_*,\theta_*,\varphi_*)}+\mathcal{O}\paren{1\over r^3}.
}
Here $(r_*,\theta_*,\varphi_*)$ is the same as Eq.~\eqref{Eq:elemag_memory_detection}.
We note that, although the effect is $r^{-2}$ order, it can be distinguished from the effect from the radiation without burst, whose order is $r^{-3}$.

\subsubsection{Coupling with photon}
As an another observable, we consider the polarization of the photon.
The relevant part of the Lagrangian is
\al{
	\mathcal{L}=-{1\over4}F_{\mu\nu}F^{\mu\nu}-\frac{c_\gamma}{4} aF_{\mu\nu}\tilde{F}^{\mu\nu}. 
}
The EoM  for the gauge field in the Lorenz gauge is
	\al{
		\p^2 A^\mu- c_\gamma \epsilon^{\mu\nu\rho\sigma} \p_\nu a \p_\rho A_\sigma=0.
	}

As a simple situation, we consider the propagation of the photon to $z$-direction, and suppose that the polarization is $x$-direction. That is,
\al{
&A_1 \propto e^{ik\cdot x},  && A_{0,2,3}=0, && k^\mu=(\omega, 0, 0, \omega).
}
are satisfied at the initial time.\footnote{This is consistent with the Lorenz gauge condition.} We expect that the axion pulse changes the polarization of the photon. By putting this background solution, we want to obtain the time evolution of $A_2$. We denote the perturbation of $A_2$ as $\delta A_2$, which follows the equation 
\al{
		\p^2 \delta A_2-c_\gamma \br{(\p_3 a)\p_0 - (\p_0 a)\p_3} A_1=0.
	}
We solve it with the large $r$ approximation. The axion field $a$ can be written as
\al{&
a=\delta(t-r) {F(\theta,\varphi)\over r},
&&
\p_3 a\simeq - \hat{x}_3\delta'(t-r) {F(\theta,\varphi)\over r},
&&
\p_0 a\simeq \delta'(t-r) {F(\theta,\varphi)\over r},
}
where $F$ is a function. From this, we obtain 
\al{
	\p^2\delta A_2 \propto -i c_\gamma \paren{1-\hat{x}_3}\delta'(u){F\over r}e^{ik\cdot x}.
}
Unfortunately, the left-hand side does not contain the term $\p_u^2\delta A_2$.   
Thus, the solution $\delta A_2$ does not have the step function term $\theta(u)$, while it has $\delta(u)$. To obtain $\theta(u)$, we have to go to $r^{-2}$ order. 


\section{Summary}\label{Sec:summary}
We have investigated various memory effects induced by the kick of the particle interacting with graviton, photon and axion. The memory effects are characterized by the terms in the radiation fields proportional to the step function, the delta function and its derivative. We have derived the memory effects by solving the classical equations of motion with the classical kick currents. We have found that the classical radiation fields up to subsubleading order can be correctly reproduced the soft factors which are consistent with known soft theorems. The axion memory starts from subleading order, which is consistent with the soft pion theorem. We have argued the decomposition of the $E$ and $B$ modes of the classical gauge field, where it has been found that the $B$-mode leading electromagnetic memory is possible if a hard particle has the magnetic charge. Then, we have investigated the detection of the memory effect. It is known that the trajectory and time delay of the test particle can be used to detect the leading and subleading  gravitational or electromagnetic memory. In addition, we have pointed out that the leading memory effect can be probed by the permanent change of the direction of the spin. The subleading axion memory can also be detected by the permanent change of the spin. Since the field value of the axion field is related to the effective theta parameter, we could detect the memory effect through the measurement of the electric dipole moment. We will leave this study for a future publication. On the other hand, the subsubleading term in gravitational field does not lead to the step function terms in the trajectory, time delay and the spin direction.

\subsection*{Acknowledgement}
We thank Alexei A. Deriglazov, Sho Higuchi, Hikaru Kawai and Koji Tsumura for helpful discussions and comments.
This work is supported in part by the Grant-in-Aid for Japan Society for the Promotion of Science Fellows No.16J06151 (YH) and No.16J01004 (SS).

\appendix
\section{The decomposition to E and B modes}\label{App:EB_decomposition}
For the $U(1)$ gauge theory, the definition of $E$, $B$ decomposition is~\cite{Winicour:2014ska,Madler:2016ggp}
\al{
X_B=D_B \Phi+\epsilon_{C B}D^C\Psi,
}
where $X_B$ is the vector on $S^2$, and $\Phi, \Psi$ are $E$-mode and $B$-modes, respectively. The $E$ and $B$ modes correspond to the parity even and odd modes, respectively.
In $(\theta,\varphi)$ or $(z,\bar{z})$ coordinates, we can write this as
\al{&
X_{\theta}=\p_\theta\Phi-{\epsilon_{\theta\varphi}\over \sin^2\theta}\p_\varphi\Psi,
&&
X_{\varphi}=\p_\varphi\Phi+{\epsilon_{\theta\varphi}}\p_\theta\Psi,
\\&
X_{z}=\p_z\Phi+i{\epsilon_{\theta\varphi}\over 2}{1+|z|^2\over|z|}\p_{z}\Psi,
&&
X_{\bar{z}}=\p_{\bar{z}}\Phi-i{\epsilon_{\theta\varphi}\over 2}{1+|z|^2\over|z|}\p_{\bar{z}}\Psi.
}
It can be seen that $\p_{\bar{z}} X_{z}-\p_{z} X_{\bar{z}}\neq0$ signals the B-mode, while the E-mode is present if $\p_{\bar{z}} X_{z}+\p_{z} X_{\bar{z}}\neq0$.\footnote{Here we have neglected the zero mode contribution.}

 Note that $(1+|z|^2)/|z|=2/\sin\theta$. 
Because of $\epsilon_{\theta\varphi}=\sin\theta$ and $\epsilon_{z\bar{z}}=-2i/(1+|z|^2)^2$~\cite{Winicour:2014ska}, we obtain
\al{
&X_\theta=\p_\theta\Phi-{1\over \sin\theta}\p_\varphi\Psi,
&&X_{\varphi}=\p_\varphi\Phi+{\sin\theta}\p_\theta\Psi,\\
&X_z=\p_z\Phi+i\p_{z}\Psi,
&&X_{\bar{z}}=\p_{\bar{z}}\Phi-i\p_{\bar{z}}\Psi.
}

Regarding the gravity, the $E$ and $B$ decomposition is given by~\cite{Nichols:2017rqr}
\al{\label{Eq:EB_decomposition_gravity}
h_{AB}={1\over2}\paren{2D_AD_B-\gamma_{AB}D^2}\Phi+\epsilon_{C(A}D_{B)}D^C\Psi,
}
in the Bondi gauge.
In $(\theta,\varphi)$ or $(z,\bar{z})$ coordinate, we can write Eq.~\eqref{Eq:EB_decomposition_gravity} explicitly:
\al{
&h_{\theta\theta}=\br{\p_\theta^2-{1\over2}{1\over\sin\theta}\p_\theta\paren{\sin\theta\p_\theta}-{1\over2}{1\over\sin^2\theta}\p_\varphi^2}\Phi-{1\over\sin\theta}\paren{\p_\theta\p_\varphi-{1\over\tan\theta}\p_\varphi}\Psi,
\nn&h_{\theta\varphi}=h_{\varphi\theta}=\paren{\p_\theta\p_\varphi-{1\over\tan\theta}\p_\varphi}\Phi+{1\over2}\br{\sin\theta\,\p_\theta^2-{1\over\sin\theta}\paren{\p_\varphi^2+\cos\theta\sin\theta \,\p_\theta}}\Psi,
\nn&h_{\varphi\varphi}=-\sin^2\theta \,h_{\theta\theta},
\nn&h_{zz}=
\paren{\p_z+{2\bar{z}\over1+|z|^2}}\p_z\Phi+i\paren{\p_z+{2\bar{z}\over1+|z|^2}}\p_z\Psi, 
\quad\quad h_{z\bar{z}}=0.
}
It can be seen that if $D_{\bar{z}}^2 h_{zz}-D_z^2 h_{\bar{z}\bar{z}}\neq0$ there is B-mode, while the E-mode is present if $D_{\bar{z}}^2 h_{zz}+D_z^2 h_{\bar{z}\bar{z}}\neq0$.

\section{Soft pion theorem for spin $1$ particle}\label{App:soft_pion_theorem_spin1}
In Ref.~\cite{Hamada:2017atr}, it is shown that the soft pion theorem for spin $0$ and spin $1/2$ external lines can be written in a universal form:
\al{\label{softNGthrm}
	\lim_{\omega \to 0} \outst a_{\omega\hat{\vb{q}}}^{(\uppi)} \, \mathcal{S} \inst
	= J^{(1)}(\vb{q}) \outst \mathcal{S} \inst,
}
where
\al{
J^{(1)}(\vb{q}) = \sum_{k} \frac{-i y \, \eta_k}{2m\, p_k\cdot q} \epsilon_{\mu\nu\rho\sigma} q^\mu p_k^{\nu} J_k^{\rho\sigma}.
}
In this appendix, we show that this formula persists even when the external particle is the spin $1$ particle. The coupling between the axion $\uppi$ and spin $1$ particle can be written as
\al{
\uppi \tr\paren{F_{\mu\nu}\tilde{F}^{\mu\nu}}=\uppi \tr\br{2\p_\mu\paren{A_\nu\p_\rho A_\sigma+{2\over3}ig A_\nu A_\rho A_\sigma}\epsilon^{\mu\nu\rho\sigma}}.
}
Note that the second term does not give rise to the pole, and therefore does not contribute to the $\mathcal{O}(1)$ soft theorem.

Let us consider the diagram where external photon line has polarization vector $e^\beta$. Then, if we attach the soft pion to this diagram, the $e^\beta$ is modified as
\al{
e^\beta \to {1\over q\cdot p}q_\mu p_\rho \epsilon^{\mu\beta\rho\alpha}e_\alpha,
}
in the soft limit. This is nothing but
\al{
{1\over q\cdot p}q_\mu p_\rho \epsilon^{\mu\gamma\rho\alpha}[J_{\alpha\gamma},e^\beta].
}
Note that the action of $J_{\mu\nu}$ is given by
\al{
[J_{\mu\nu},A_\rho]=i\paren{\eta_{\nu\rho}A_\mu-\eta_{\mu\rho}A_\nu}.
}
Therefore, we confirm that the formula in Ref.~\cite{Hamada:2017atr} holds for spin $1$ particle.

One might think that the formula can be extended to the spin $2$ particle, where the interaction is given by $a\tilde{R}_{\mu\nu}R^{\mu\nu}$. However, this contain at least four derivative, and does not contribute to the subleading soft theorem. In this sense, Eq.~\eqref{softNGthrm} is not valid for spin $2$ particle.

\bibliographystyle{TitleAndArxiv}
\bibliography{Bibliography}

\end{document}